# A structural perspective on the origin of the anomalous weak-field piezoelectric response at the polymorphic phase boundaries of (Ba, Ca)(Ti, M)$O_3$ lead-free piezoelectrics

Mulualem Abebe[1], Kumar Brajesh[1], Anupam Mishra[1], Anatoliy Senyshyn[2] and Rajeev Ranjan[1]*

[1]Department of Materials Engineering, Indian Institute of Science Bangalore-560012, India

[2]Forschungsneutronenquelle Heinz Maier-Leibnitz (FRM II). Technische Universität München, Lichtenbergestrasse 1, D-85747 Garching b. München, Germany

## Abstract

Although, as part of a general phenomenon, the piezoelectric response of $Ba(Ti_{1-y} M_y)O_3$ (M= Zr, Sn, Hf) increases in the vicinity of the orthorhombic (*Amm*2) –tetragonal (*P*4*mm*) and orthorhombic (*Amm*2)-rhombohedral (*R*3*m*) polymorphic phase boundaries, experiments in the last few years have shown that the same phase boundaries show significantly enhanced weak-field piezo-properties in the Ca-modified variants of these ferroelectric alloys, i.e., (Ba,Ca)(Ti, M)$O_3$. So far there is a lack of clarity with regard to the unique feature(s) which Ca modification brings about that enables this significant enhancement. Here, we examine this issue from a structural standpoint with M = Sn as a case study. We carried out a comprehensive comparative structural, ferroelectric and piezoelectric analysis of the *Amm*2 phase in the immediate vicinity of the *P*4*mm* - *Amm*2 phase boundaries of (i) Ca-modified Ba(Ti,Sn)$O_3$, as per the nominal formula $(1-x)BaTi_{0.88}Sn_{0.12}O_3-(x)Ba_{0.7}Ca_{0.3}TiO_3$ and (ii) without Ca modification, i.e. $Ba(Ti_{1-y}Sn_y)O_3$. We found that the spontaneous lattice strain of the *Amm*2 phase is noticieably smaller in the Ca-modified counterpart. Interestingly, this happens alongwith an improved spontaneous polarization by enhancing the covalent character of the Ti-O bond. Our study suggests that the unique role of Ca-modification lies in its ability to induce these seemingly contrasting features (reduction in spontaneous lattice strain but increase the polarization).

*ranjanrajeeb@gmail.com

## I. Introduction

Environmental concerns have triggered a surge in research activities in lead-free piezoelectrics. Guided primarily by the important role of the morphotropic phase boundary (MPB) in enhancing the piezoelectric properties of lead-based ferroelectric alloys [1], the search for equivalents in the lead-free category has been directed towards compositional design that can induce interferroelectric instability leading to morphotropic/polymorphic phase boundaries near room temperature. From the phenomenological viewpoint, the large piezoelectric response at an interferroeelctric instability is associated with a reduction in the crystalline anisotropy of polarization, leading either to rotation of the polarization vector within the unit cell and/or to significant increase in the domain wall motion [2-6]. Among the lead-free piezoelectrics $BaTiO_3$-based alloys have shown the highest weak-field piezoelectric coefficient $d_{33}$ exceeding even the commercially available soft-PZT [7-11]. The strategy to increase the piezo-response of $BaTiO_3$ at ambient conditions rests on exploiting the interferroelectric instabilities inherent in $BaTiO_3$. $BaTiO_3$ shows three structural transformations as a function of temperature: (i) cubic (Pm3m) - tetragonal (P4mm) at 130 $^oC$, (ii) tetragonal (P4mm) – orthorhombic (Amm2) at ~ 0 $^oC$, and (iii) orthorhombic (Amm2) - rhombohedral (R3m) at -90 $^oC$ [1]. The two latter transformations P4mm-Amm2 and Amm2-R3m are inter-ferroelectric in nature and are important from the viewpoint of enhancing the piezoelectric response. Chemical modifications which can push the P4mm-Amm2 and the Amm2-R3m transitions from below room temperature to room temperature and above are expected to increase the room temperature piezoelectric and dielectric response of the system. Although, the piezoelectric properties of modified $BaTiO_3$ were not seriously pursued before this decade, it was known that substitution of Zr, Sn and Hf for Ti in $BaTiO_3$ increases the P4mm-Amm2 and the Amm2-R3m transition temperatures [12-17]. Subsequent studies confirmed this phenomena and also showed merging of the three phase boundaries [18-20], just above room temperature. Recently, Ce modification of $BaTiO_3$ has also been reported to bring about a similar change [21]. The enhancement of the piezoresponse in these systems has also been reported in the vicinity of their respective phase boundaries[21-25].

A major breakthrough was the discovery by Liu and Ren of anomalously large weak-field piezoelectric response in (Ba,Ca)(Ti, Zr)$O_3$ (BCTZ) with $d_{33}$ > 500 pC/N [7]. Sooner, the anomalous piezoresponse was also reported in (Ba,Ca)(Ti, Sn)$O_3$ (BCTS) [8-10] and (Ba,Ca)(Ti, Hf)$O_3$ (BCTH) [11]. Liu and Ren attributed the anomalous piezoelectricity to the proximity of the polymorphic phase (denoted as P4mm-R3m in their work [7]) boundary to the *P4mm-R3m-Pm3m* triple point which was also suggested to be tricritical in nature[7]. Later studies, however, confirmed the existence of a narrow orthorhombic Amm2 phase region between the *P4mm* and *R3m* phase regions[26, 27], and thereby established that the sequence of phase transitions in BCTZ to be similar to that of the parent compound $BaTiO_3$ and Ba(Ti, M)$O_3$ (M = Zr, Sn, Hf) [18-20]. The region in the phase diagram where the three phase boundaries *Pm3m-P4mm, P4mm-Amm2, Amm2-R3m* converge has been labelled as convergence region by Keeble et al [26]. Since thermodynamic arguments preclude coexistence of four phases in equilibrium for a pseudo-binary phase diagram, the issue was resolved in a computational study which predicted two triple points in close proximity [28]. Detailed experiments have shown that $d_{33}$ exhibits maximum at the polymorphic phase boundary away from the convergence region [29]. Acosta et al have argued that although the reduction in polarization anisotropy is an important consideration, it is not a sufficient condition for enhancement of the piezoelectric properties [30]. Apart from the phenomenological thermodynamic explanations, attempts have also been made to understand the enhanced properties in these systems in terms of increased propensity of field driven structural transformation [31, 32] and enhancement in the domain wall motion as the phase boundary is approached [33]. Such trends are, however, common across different MPB systems, and do not specifically address the anomalous $d_{33}$ observed in the Ca-modified variants of Ba(Ti, Me)$O_3$ vis-à-vis the Ca-free counterparts.

Eventhough, as argued by Liu and Ren [7], a drastic decrease in the polarization anisotropy may be achieved as the convergence region is approached, the polarization is also expected to decay because of the system's proximity to the cubic phase. This will have a detrimental effect on the piezoelectric response [30]. The best piezo-response is therefore expected when the system is optimally away from the convergence region. *The question therefore is: is such an optimal situation accessible only in the Ca-modified variants of Ba(Ti, Zr)$O_3$, Ba(Ti, Sn)$O_3$ and Ba(Ti, Hf)$O_3$ and not in the Ca-free counterparts?* It may be difficult to answer this precisely using the phenomenological free energy calculation approach, as the energy difference may not be significantly enough for an unambiguous comparison. An alternative approach would be to identify a distinct structural / microstructural feature brought in by Ca-substitution and relate it to the enhanced piezoelectric response. With this as motivation, we carried out a comparative structural analysis of Ca-substituted and Ca-free Ba(Ti, Sn)$O_3$ as a case study. For the Ca-modified system we investigated the series (1-x)Ba($Ti_{0.88}Sn_{0.12}$)$O_3$ – (x)$Ba_{0.7}Ca_{0.3}TiO_3$, first reported by Xue et al [8]. Our systematic structural analysis of the compositions near the polymorphic phase boundaries of the two series revealed a distinctly reduced spontaneous lattice strain, both in the orthorhombic and the tetragonal phases of the Ca-modified variant. We also found the Ti-O bond to be more covalent in character in the Ca-modified variant which improves the polarization. We associate these subtly distinct features as contributing to the anomalous increase in the weak-field piezoresponse of the Ca-modified variants.

## II. Experimental

Ceramic specimens of (1-x)Ba($Ti_{0.88}Sn_{0.12}$)$O_3$ –(x)$Ba_{0.7}Ca_{0.3}TiO_3$ (to be denoted in short as BCTS) and Ba($Ti_{1-y}Sn_y$)$O_3$ (to be denoted as BTS) were prepared via conventional solid-state route. High purity $BaCO_3$ (99.8%; Alfa Aesar), $CaCO_3$ (99.99%; Alfa Aesar), $TiO_2$ (99.8%; Alfa Aesar) and $SnO_2$ (99.99%; Alfa Aesar) were thoroughly mixed in zirconia jars with zirconia balls and acetone as the mixing medium using a planetary ball mill (Fritsch P5). The thoroughly mixed powder was calcined at $1300^0$C for 4 hrs and milled again in acetone for 5hrs for better homogenization. The calcined powder was mixed with 2 wt. % polyvinyl alcohol (PVA) and pressed into disks of 15 mm diameter by using uniaxial dry pressing at 10 ton. Sintering of the pellets were carried out at $1550^0$C for 4 hours under ambient condition. X-ray powder diffraction was done using a Rigaku (smart lab) with Johanson monochromator in the incident beam to remove the Cu-K$\alpha_2$ radiation. Neutron powder diffraction data was collected on the SPODI diffractometer at FRM II, Germany. Microstructure of the sintered pellets was recorded by Scanning Electron Microscopy (ESEM, Quanta) after gold sputtering the polished and thermally etched surface. Dielectric measurement was carried out using Novocontrol (Alpha AN) impedance analyzer. Measurement of the direct weak-field longitudinal piezoelectric coefficient ($d_{33}$) was carried using piezotest PM 300 by poling the pellets at room temperature for 1h at a field of ~2.2 kV/mm. Strain loop and polarization electric field hysteresis loop were measured with a Precision premier II loop tracer. Structure refinement was carried out using FULLPROF software [34].

## III. Results

### A. Ferroelectric, piezoelectric and structural evolution

Fig. 1 shows composition variation of the weak-field direct longitudinal piezoelectric coefficient ($d_{33}$), spontaneous and remanent polarization of BCTS at room temperature. The sharp increase in $d_{33}$ with increasing x in the composition range $0<x<0.13$ mimics the trends in the saturation and remanent polarizations. The rate of increase of $d_{33}$ with composition becomes less in the range $0.13<x<0.21$, reaching a maximum of 475 pC/N at x= 0.21. For x > 0.21, $d_{33}$ drops slightly. The saturation and the remanent polarizations also show maxima at x =0.21 which, as will shown below, correspond to the orthorhombic (*Amm2*) phase adjacent ot the *P4mm –Amm2* boundary. Fig. 2 shows the compositional evolution of representative pseudocubic x-ray Bragg profiles. Consistent with the anomalous changes in the properties at x = 0.13 and x =0.21, an abrupt change in the Bragg profiles can be seen at this two boundary composiitons. For example, for $x < 0.13$, the $\{400\}_{pc}$ is a singlet in conformity with rhombohedral structure. This shape abruptly changes at x =0.13. The splitting of the $\{400\}_{pc}$ profile, with nearly equal intensity for x =0.13 is consistent with orthorhombic (Amm2) distortion of the perovskite cell. The Bragg profiles remain nearly unchanged in the composition range $0.13<x<0.21$. This is consistent with the nearly constant $d_{33}$ in the same composition range, Fig 1a. For x ≥ 0.22, the nature of the $\{400\}_{pc}$ splitting changes abruptly- not only the separation between the peaks increases, but also the intensity ratio is close to 1:2. This suggests onset of a tetragonal (P4mm) distortion. This transformation is accompanied by a slight decrease in $d_{33}$ at x =0.21, Fig1a. Rietveld analysis of the XRD patterns, shown in Fig. 3, confirmed a single phase Amm2 for $0.13<x<0.21$. However, the XRD pattern of x = 0.22 could be satisfactorily fitted *P4mm* + *Amm2* structural model, Fig. 3d. The fraction of the Amm2 phase decreases for x > 0.22.

### B. Dielectric study and Phase diagram

Fig 4 shows the real part of the relative permittivity of BCTS as a function of temperature measured at 1kHz. The tetragonal composition x =0.29 shows one dielectric anomaly at 72$^{o}$C corresponding to a paraelectric-ferroelectric (*P3mm-P4mm*) transformation. The anomaly temperature decreases with decreasing x. For x = 0.13, the Curie point decreases to 50 $^{o}$C. At the same time another weak anomaly becomes visible at 35$^{o}$C. Since the structure at room temperature (27 $^{o}$C) for this composition is orthorhombic Amm2, the dielectric anomaly at 35$^{o}$C can be attributed to the *Amm2-P4mm* transformation. This composition is also expected to show orthorhombic (*Amm2*)- rhombohedral (*R3m*) transformation below the *Amm2-P4mm* transition. However, the corresponding dielectric anomaly is not visible in the real part of the permittivity plot. We looked for its signature in the imaginary part of permittivity-temperature data, Fig. 5. The three anomalies indeed become visible for compositions in the range $0.13<x<0.21$in the imaginary part of the permittivity data - the third anomaly corresponding to Amm2-R3m interferroelectric transformation occurs just below room temperature. Fig. 6(a) depicts a phase diagram drawn on the basis of the dielectric anomalies observed in the imaginary part of the permittivity. From this diagram it is evident that the three anomalies tend to merge somewhere close to x = 0.10, T = 40 $^{o}$C. A characteristic feature of the composition close to the convergence region is that it shows very large value of the permittivity maximum as compared to other compositions, Fig. 4.

### C. Evidence of relaxor ferroelectricity

Fig. 7 (a,b) shows the temperature dependence of relative permittivity of x = 0.13 at different frequencies. The permittivity at the different frequencies branches off below 100 $^{o}$C, both in the real and imaginary parts. A Curie-Weiss analysis of the $\varepsilon'(T)$ data reveals a departure below 100 $^{o}$C, Fig. 7c. We also carried out high temperature x-ray diffraction study of this composition and found the cubic lattice parameter to deviate from the linear trend below 100 $^{o}$C, Fig. 7d. The dielectric maximum temperature also shows a slight increase with increasing frequency. All these features are in conformity with the relaxor ferroelectric behaviour of the system in the temperature range 50- 100 $^{o}$C [35, 36]. The Burns temperature appears to be 100 $^{0}$C below which dynamic polar nano regions are likely to appear. A Vogel-Fulcher analysis of the frequency dependent permittivity maximum temperature gave the freezing temperature ($T_f$) of the polar nano regions as 48.8 $^{o}$C. The dielectric dispersion, however, persists below $T_f$ and becomes pronounce in the vicinity of *P*4*mm*-*Amm*2 and *Amm*2-*R*3*m* interferroelectric transitions. From these observations, we may infer that the ferroelectric state at room temperature is most likely to be a mixture of long and short ranged polar order. A similar behaviour was also reported recently for morphotropic phase boundary composition of $(Ba,Ca)(Ti, Zr)O_3$ [31].

### D. Comparison with $Ba(Ti, Sn)O_3$

For sake of comparison we also show in Fig. 6b the phase diagram of the Ca-free system $Ba(Ti_{1-y}Sn_y)O_3$ (BTS) drawn using the dielectric anomalies (not shown). Qualitatively, this phase diagram is similar to that of BCTS (Fig. 6a), with the three phase boundaries *Pm*3*m*-*P*4*mm*, *P*4*mm*-*Amm*2 and *Amm*2-*R*3*m* merging at ~ 11 mole percent of Sn at ~ 45 $^{o}$C. As per this phase diagram, and also reported earlier [20, 37, 38], the *P4mm-Amm2* and *Amm2-R3m* boundaries at room temperature are at y = 0.05 and y = 0.08, respectively. Amm2 phase is stable in the composition range $0.05 \le y \le 0.08$. The corresponding composition range of the *Amm2* phase in BCTS is $0.13 \le x \le 0.21$. Evidently x = 0.21 in the Ca-modified series (BCTS-21) and y=0.05 in the Ca-free series (BTS-05) are analogous to each other as they represent *Amm*2 phase in the immediate vicinity of the *P*4*mm*-*Amm*2 boundary of their respective phase diagrams. Accordingly, we selected these two compositions (BCTS-21 and BTS-05) for a detailed structural comparison. The $d_{33}$ of BTS-05 and BCTS-21 was measured as 292 and 475 pC/N, respectively. The polarization and strain – field hysteresis loops of the two compositions are shown in Fig. 8. Both the maximum strain and saturation polarization is higher for BCTS-21 than BST-05. On careful examination we also found the coercive field of BCTS-21 to be slightly lower and the large-field electro-strain relatively higher as compared to that of BTS-05 (Fig. 8). These features are indicative of a relatively better domain wall mobility in the Ca-modified variant (BCTS-21) as compared to its Ca-free counterpart (BST-05). Keeping in view the fact that piezoelectric and ferroelectric properties in ceramics are also dependent on grain size (in general, piezoceramics with small grains show lower polarization and $d_{33}$ [43]), it is important to rule out if the lower polarization and $d_{33}$ in the Ca-free variant is due to its relatively smaller grains or not. Fig. 9 shows that average grain size of BTS-05 is nearly five times larger (~80 mirons) than that of BCTS-21 (~ 18 microns). This confirms that the substantially lower $d_{33}$ of BTS-05 as compared to BCTTS-21 cannot be a grain size effect. We are therefore justified in seeking the origin of the anomalous difference in the property in the structural difference between the orthorhombic phases of the two variants.

Fig. 10 compares some representative pseudocubic x-ray Bragg profiles of both BTS-05 and BCTS-21. The most revealing contrast can be seen in the doublet-like nature of the $\{400\}_{pc}$ profile – the split peaks are more distant in BTS-05 than in BCTS-21 suggesting the degree of orthorhombic distortion to be more in the former. Although, due to slightly better resolution of our x-ray diffractometer, the XRD data gives us accurate information regarding the nature and magntitude of the lattice distortion, it cannot yield precise estimation of the cation-oxygen bond-distances because of (i) poor scattering amplitude of oxygen for x-rays (x-ray scattering factor is proportional to the atomic number) and (ii) the form factor dependence of the scattered intensity which makes structural analysis not very meaningful using the weak Bragg peaks at high angles. In contrast, neutrons can be as efficiently scattered by low Z atoms (such as oxygen in our case) as with atoms with high Z values, and its (nuclear) scattering length has no form factor dependence. This makes neutron powder diffraction an ideal tool for precise determination of oxygen coordinates in perovskite structures which are only slightly deviated from the most symmetric cubic form [39 - 42]. In our particular case, the negative coherent scattering length of Ti (b = -3.44 fm) as compared to that of oxygen ( b = 5.803 fm) makes the scattering contrast significantly large between Ti and O. As a consequence, the Ti-O distances can be very accurately ascertained by Rietveld analysis of neutron diffraction data. In conformity with the XRD analysis presented above (Fig. 3), the neutron powder diffraction patterns of BST-05 and BST-21 were nicely fit with a single phase *Amm*2 structural model. The Rietveld fitted neutron diffraction patterns are shown in Figure 11. Similar to the case with the XRD the neutron Bragg profiles of BST-05 also shows noticeably larger splitting as compared to that in BCTS-21 (insets of Fig. 11), confirming larger magnitude of orthorhombic lattice distortion in the Ca-free variant. The refined structural paramerters and the Ti-O bond lengths are given in Table 1.

## IV. Discussion

We may point out that in contrast to Ref [8], wherein $d_{33}$ is shown to exhibit maximum at x=0.30, in our samples we found the maximum to occur at x = 0.21. This difference of 9 mole percent in BTS concentration actually corresponds to a difference of only 2 mole percent in terms of Ca concentration (9 mole percent Ca in our case as compared to 7 mole percent Ca in Ref [8]). The difference can be attributed to the slight difference in the raw materials used by the different groups and slanted nature of the phase boundary near room temperature which makes the property a bit insensitive to the slight inadvertent change in composition. We made samples in different batches and found reporoducibility of the results for any given composition in our series. Importantly, however, the slight difference/shift in the composition reported by us and in Ref [8] is not expected to influence the problem addressed in our paper which relies primarily on the comparative analysis of the similar looking structures near the polymorphic phase boundaries of $Ba(Ti_{1-y}Sn_y)O_3$ and (1-x)$Ba(Ti_{0.88}Sn_{0.12})O_3$ – x$(Ba_{0.7}Ca_{0.3})TiO_3$ on specimens prepared using the same raw materials.

Within the framework of phase digram of the analogous system $(Ba,Ca)(Ti,Zr)O_3$, Keeble *et. al.* argued that the large piezoelectric response is associated with the increased instability gradient of the intermediate orthorhombic phase [26]. The authors pointed out that it requires a comparatively less Zr fraction in BCTZ (~13 mole percent of Zr) as compared to that in BTZ (~15 mole percent of Zr) to arrive at the convergence region. The suppression of the orthorhombic phase at a slightly lower Zr concentration in BCTZ was attributed to the setting in of octahedral tilt by Ca-substitution ($CaTiO_3$ shows tilted octahedra structure [39]). Similar to BCTZ, the convergence region in our BCTS occurs at lower Sn concentrations (9 mole percent) than in BTS (11 mole percents of Sn). While an unambiguous explanation is difficult to arrive at by mere comparison of the two phase diagrams, we do find a distinctive difference in the magnitude of the sponataneous lattice disitortion in BCTS and BTS in the immediate vicinity of their respective *P*4*mm*-*Amm*2 phase boundaries. The spontaneous lattice strain is a measure of deviation of the lattice from a cubic-like lattice. In diffraction patterns this manifests as splitting of the pseudocubic Bragg reflections. As evident from Fig.10, the two split reflections in the pseudocubic $\{400\}_{pc}$ x-ray Bragg profile are significantly more separated in BTS-05 than in BCTS-21. This suggests a larger spontaneous lattice strain in BTS-05. In analogy with the tetragonal (*P*4*mm*) phase, the splitting in the pseudocubic $\{h00\}_{pc}$ Bragg profiles of which is manifestation of the spontaneous tetragonal strain ($c/a$ -1), we define a "pseudo-tetragonlity" parameter for the orthorhombic *Amm*2 phase. In this regard, we first convert the orthorhombic lattice parameters ($a, b, c$) to a corresponding pseudocubic lattice parameters ($a_{pc}, b_{pc}, c_{pc}$) using the approximate relationships: $a_{pc} = a$, $b_{pc} = b/\sqrt{2}$, $c_{pc} = c/\sqrt{2}$. The equality $a_{pc}, b_{pc}, c_{pc}$ would imply a strain-free cubic lattice. Using the refined *Amm*2 lattice parameters given Table 1 for both BTS-05 and BCTS-21 we found that $b_{pc}$ and $c_{pc}$ values are close to each other, and that both are distinctly larger than $a_{pc}$. We therefore define an average $c_{av}$ as $(b_{pc} + c_{pc})/2$ and subsequently define a "pseudo-tetragonality" of the orthorhombic (*Amm*2) phase as $\eta_o = c_{av}/a_{pc} - 1$. $\eta_o$(BTS-05) = 0.004 and $\eta_o$(BCTS-21) = 0.002. The spontaneous lattice strain of the Ca-free variant is nearly double the value of Ca-modified variant. This difference can as well be seen in the neutron powder diffraction patterns of BTS-05 and BCTS-21. The high angle pseudocubic neutron Bragg profiles shows clear splitting in BTS-05, whereas it is not case in BCTS-21. We also found that the spontaneous tetragonal lattice strain defined as $\eta_T = c_T/a_T - 1$, where $a_T$ and $c_T$ are tetragonal lattice parameters, follows the same trend for the Ca-free and Ca-modified variants in the immediate vicinity of the *P*4*mm*-*Amm*2 boundaries. A similar scenario can be found by comparing the lattice parameters of the orthorhombic phase of (Ba, Ca)(Ti, Zr)$O_3$ [32] and its Ca-free counterpart Ba(Ti,Zr)$O_3$ [23]. Recently, Tutuncu et al have demonstrated that the domain wall motion is enhanced with decreasing tetragonal spontaneous lattice strain as the MPB composition is approached in the analogous system (Ba, Ca)(Ti, Zr)$O_3$ [33]. While this scenario is independently true for any ferroelectric alloy in the vicinity of MPB, what is of significance in our context is that the spontaneous lattice strain is noticeably less in the Ca-modified variant when compared with its Ca-free counterpart. We can therefore expect that the domain wall motion to be relativey more in the Ca-substituted system. The larger electrostrain in BCTS-21 as compared to BTS-05, shown in Fig. 8, is consistent with our viewpoint.

The other important factor is improvement of polarization in the Ca-modified variant. First principles studies have shown that spontaneous polarization in $BaTiO_3$ is primarily determined by the covalent character of the Ti-O bond [44]. A higher degree of covalent character would manifest as decrease in the Ti-O distance. The six Ti-O bonds are depicted in the Amm2 cell in Fig. 12. We found that the shortest Ti-O distance in BTS-05 is 1.933 Å, whereas it is 1.910 Å in BCTS-21. A similar scenario was found in the tetragonal phases of the two variants. These results confirm that the degree of covalent character of the Ti-O bonds in Ca-substituted composition is relatively increased This is consistent with the observation of higher saturation polarization in the Ca-modified variant substituted composition, Fig. 8. Levin et al have reported that substitution of Ca at the Ba-site in $BaTiO_3$ leads to amplification of the Ti off-centering [45]. This highly anisotropic distortion appears to be responsible for sustaining /improving the polarization in the Ca-modified variants evenwhile the overall spontaneous strain is reduced [45].

## V. Conclusions

In conclusion, we carried out a comparative structural analysis of the phases of Ba(Ti, Sn)$O_3$ and its Ca-substituted counterpart (Ba,Ca)(Ti,Sn)$O_3$ in the immediate of their respective *P4mm-Amm2* boundaries to understand the factors which lead to a distinct enhancement in the piezoelectric response of the Ca-modifed variant. We found that the Ca-modified variant exhibit reduced spontaneous lattice strain as compared to its Ca-free counterpart. We also found the Ca modified variant to exhibit improved polarization by increasing the degree of covalent character of the Ti-O bond. Our results suggest that the large piezoelectric response of the Ca-modified variants of Ba(Ti,M)$O_3$ is associated with a unique combination of (i) enhanced domain wall motion enabled by reduced spontaneous lattice strain and (ii) improved polarization.

**Acknowledgement**: R. Ranjan is grateful to the Science and Engineering Research Board (SERB) of the Department of Science and Technology, Govt. of India for financial assistance (SERB/F/5046/2013-14). Kumar Brajesh gratefully acknowledges SERB for the award of National Post Doctoral Fellowship.

**References**:

[1] B. Jaffee, W. R. Cook, H. Jaffee, Piezoelectric ceramics, Academic press, London (1971).

[2] H. Fu and R.E. Cohen, Nature **403**, 281 (2000).

[3]R. Guo, L.E. Cross, S.-E. Park, B. Noheda, D.E. Cox, and G. Shirane, Phys. Rev. Lett. **84**, 5423 (2000).

[4] D. Vanderbilt and M.H. Cohen, Phys. Rev. B **63**, 094108 (2001).

[5] D. Damjanovic J Am Ceram Soc **88**, 2663 (2005).

[6] Y.M. Jin, Y.U. Wang, A.G. Khachaturyan, J.F. Li, and D. Viehland, Phys. Rev. Lett. **91**, 197601 (2003).

[7] W. Liu and X. Ren, Phys. Rev. Lett. **103**, 257602 (2009)

[8] ]D. Xue, Y. Zhou, H. Bao, J. Gao, C. Zhou, and X. Ren, Appl. Phys. Lett. **99**, 122901 (2011)

[9] Y. G. Yao, C. Zhao, D. Lv, D. Wang, H. Wu, Y. yang, and X. Ren, EPL **98** 27008 (2012)

[10] L.-F. Zhu, B.-P. Zhaang, X.-K. Zhao, L. Zhao, F.-Z. Yao, X. Han, P.-F Zhou, and J.-F. Li, Appl. Phys. Lett. **103**, 072905 (2013).

[11] C. Zhou, W. F. Liu, D. Z. Xue, X. B. Ren, H. X. Bao, J. H. Gao, and L. X. Zhan, Appl. Phys. Lett. **100**, 222910 (2012).

[12] G. H. Jonker, and W. Kwestroo, *J. Am. Ceram. Soc.* **41,** 390 (1958).

[13] T. N. Verbitskaya, G. S. Zhdanov, Yu. N. Venevtsev, S. P. Solov'ev, Kristallografiya, 3, 186; Soviet Phys. -Cryst. **3,** 182, 1958.

[14]G. H. Jonker, Philips Tech. Rev, **17**, 129 (1955).

[15] G. A. Smolenskii, V. A. Isupov, Dokl. Akad. Nauk SSSR, **96**, 53 (1954).

[16] E. G. Fesenko, O. I. Prokopalo, Soviet Phys. Cryst. **6**, 373 (1961)

[17] T. N. Verbitskaya, E. I. Gindin, V. G. Prokhvatilov, Fiz. Tverd. Tela, Sbornik, **1**, 180, (1959).

[18] V. V. Lemanov, Ferroelectrics, **354**, 69 (2007)

[19] Z. Yu, R. Guo, A. S. Bhalla, J. App. Phys. **88**, 410 (2000)

[20] A. K. Kalyani, K. Brajesh, A. Senyshyn, and R. Ranjan, Appl. Phys. Lett. **104**, 252906 (2014).

[21] K. Brajesh, A. K.Kalyani, R. Ranjan, Appl. Phys. Lett.**106**, 012907 (2014).

[22] Z. Yu, C. Ang, R. Guo, A. S. Bhalla, J. App. Phys. **92**, 1489 (2002)

[23] A. K. Kalyani, A. Senyshyn, and R. Ranjan, J. Appl. Phys. **114**, 014102 (2013).

[24] A. K. Kalyani and R. Ranjan, J. Phys. Condens. Matter **25**, 362203 (2013).

[25] A. K. Kalyani, H. Krishnan, A. Sen, A. Senyshyn, and R. Ranjan, Phys. Rev. B **91**, 024101 (2015).

[26] D. S. Keeble, F. Benabdallah, P. A. Thomas, M. Maglione, and J. Kreisel, Appl. Phys. Lett. **102**, 092903 (2013)

[27] Y. Tian, X. Chao, L. Jin, L. Wei, P. Lianag, and Z. Yang, Appl. Phys. Lett. **104**, 112901 (2014).

[28] A. A. Heitmann, and G. A. Rossetti, Jr. J. Am. Ceram. Soc. **97**, 1661 (2014).

[29] M. Acosta, N. Novak, W. Jo, and J. Roedel, Acta Mater. **80**, 48 (2014)

[30] M. Acosta, N. Khakpash, T. Someya, N. Novak, W. Jo, H. Nagata, G. A. Rossetti, Jr., J. Roedel, Phys. Rev. B. **91**, 104108 (2015)

[31] K. Brajesh, K. Tanwar, M. Abebe and R. Ranjan, Phys. Rev. B **92**, 224112 (2015).

[32] K. Brajesh, M. Abebe and R. Ranjan, Phys. Rev. B **94**, 104108 (2016).

[33] G. Tutuncu, B. Li, K. Bowman, and J. L. Jones, J. Appl. Phys. **115**, 144104 (2014).

[34] Rodrigues-J. Carvajal. FullPROF 2000 A Rietveld Refinement and Pattern Matching Analysis Program. France: Laboratories Leon Brillouin (CEA-CNRS)

[35] A. A. Bokov, and Z. –G. Ye, Phys. Rev. B **65**, 144112 (2002)

[36] J. Kreisel, P. Bouvier, M. Maglione, B. Dkhil, A. Simon, Phys. Rev. B. **69**, 092104 (2004)

[37] L. Veselinovic, M. Mitric, M. Avdeev, S. Markovic, and D. Uskokovic, J Appl. Cryst. **49**, 1726 (2016).

[38] G. Singh, V. S. Tiwari and P. K. Gupta, J. Appl. Cryst. **46**, 324 (2013)

[39] R. Ranjan, A. Agrawal, A. Senyshyn, and H. Boysen, J. Phys.: Condens. Matter **18**, L515 (2006)

[40] R. Garg, A. Senyshyn and R. Ranjan, J. Phys,: Condens. Matter **24**, 455902 (2012)

[41] D. K. Khatua, A. Senyshyn and R. Ranjan, Phys. Rev. B **93**, 134106 (2016)

[42] B. N. Rao, R. Datta, S. S. Chandrasekaran, D. K. Mishra, V. Sathe, A. Senyshyn and R. Ranjan, Phys. Rev. B **88**, 224103 (2013)

[43] D. Damjanovic, Rep. Prog. Phys. **61**, 1267 (1998).

[44] R. E. Cohen, Nature, **358,** 136 (1992).

[45] I. Levin, V. Krayzman, J. C. Woicik, Appl. Phys. Lett. **102**, 162906 (2013).

Table1 (a) : Refined orthorhombic structural parameters and Ti-O bond lengths of $BaTi_{0.95}Sn_{0.05}O_3$ (BTS-05)

| Space group: Amm2 | | | | |
|---|---|---|---|---|
| Atoms | x | y | z | B($Å^2$) |
| Ba/Ca | 0.000 | 0.000 | 0.000 | 0.15 (6) |
| Ti/Sn | 0.500 | 0.000 | 0.498 (4) | 0.88 (9) |
| O1 | 0.000 | 0.000 | 0.494(3) | 0.54(1) |
| O2 | 0.500 | 0.257 (9) | 0.236 (1) | 0.61 (7) |
| a= 4.0022(8) Å, b= 5.6768(1) Å, c= 5.6875(1) Å , V = 129.22(5) $Å^3$, $R_p$: 8.32, $R_{wp}$: 11.5, $R_{exp}$: 10.47, $Chi^2$: 1.20 | | | | |

| Bond | Bond length (Å) | No. of bonds |
|---|---|---|
| $Ti – O_1$ | 2.0012 | 2 |
| $Ti – O_2$ | 2.085 | 2 |
| $Ti – O_2$ | 1.933 | 2 |

Table 1(b) : Refined orthorhombic structural parameters and Ti-O bond lengths of $0.79Ba(Ti_{0.88}Sn_{0.12})O_3$ – $0.21(Ba_{0.7}Ca_{0.3})TiO_3$ (BCTS-21)

| Space group: Amm2 | | | | |
|---|---|---|---|---|
| Atoms | X | y | z | B($Å^2$) |
| Ba/Ca | 0.000 | 0.000 | 0.000 | 0.50 (2) |
| Ti/Sn | 0.500 | 0.000 | 0.513(3) | 0.65 (8) |
| O1 | 0.000 | 0.000 | 0.493 (2) | 0.61(8) |
| O2 | 0.500 | 0.251 (6) | 0.240 (1) | 0.64 (4) |
| a= 4.0035(6) Å, b= 5.6702(2) Å, c= 5.6730(3) Å , V = 128.78(9) $Å^3$, $R_p$: 2.91, $R_{wp}$: 4.07, $R_{exp}$: 1.93, $Chi^2$: 4.46 | | | | |

| Bond | Bond length (Å) | No. of bonds |
|---|---|---|
| $Ti – O_1$ | 2.0050 | 2 |
| $Ti – O_2$ | 2.104 | 2 |
| $Ti – O_2$ | 1.910 | 2 |

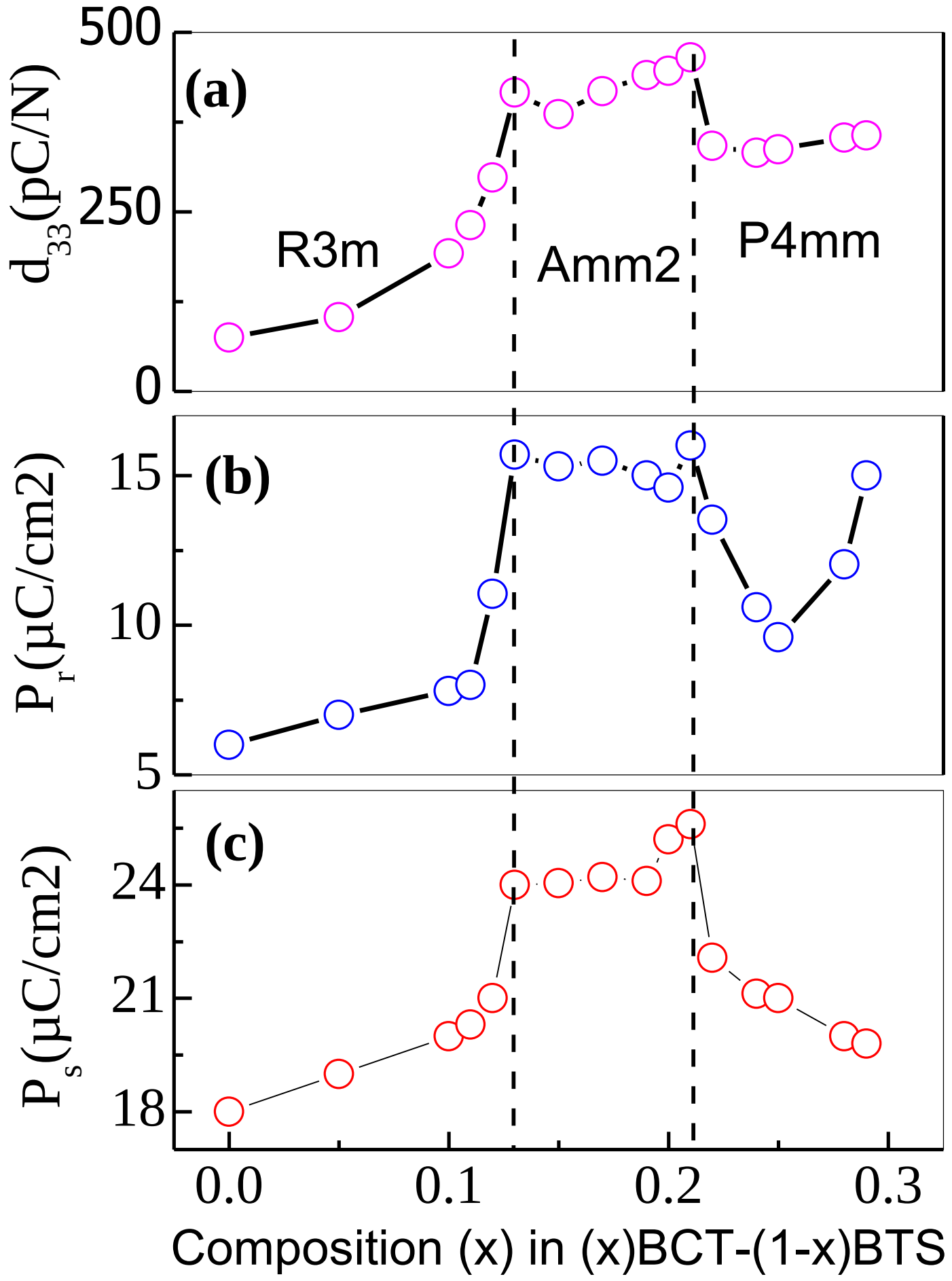

Fig 1 Composition dependence of (a) piezoelectric coefficient ($d_{33}$), (b)remnant polarization, and (c) saturated polarization of BCTS at room temperature.

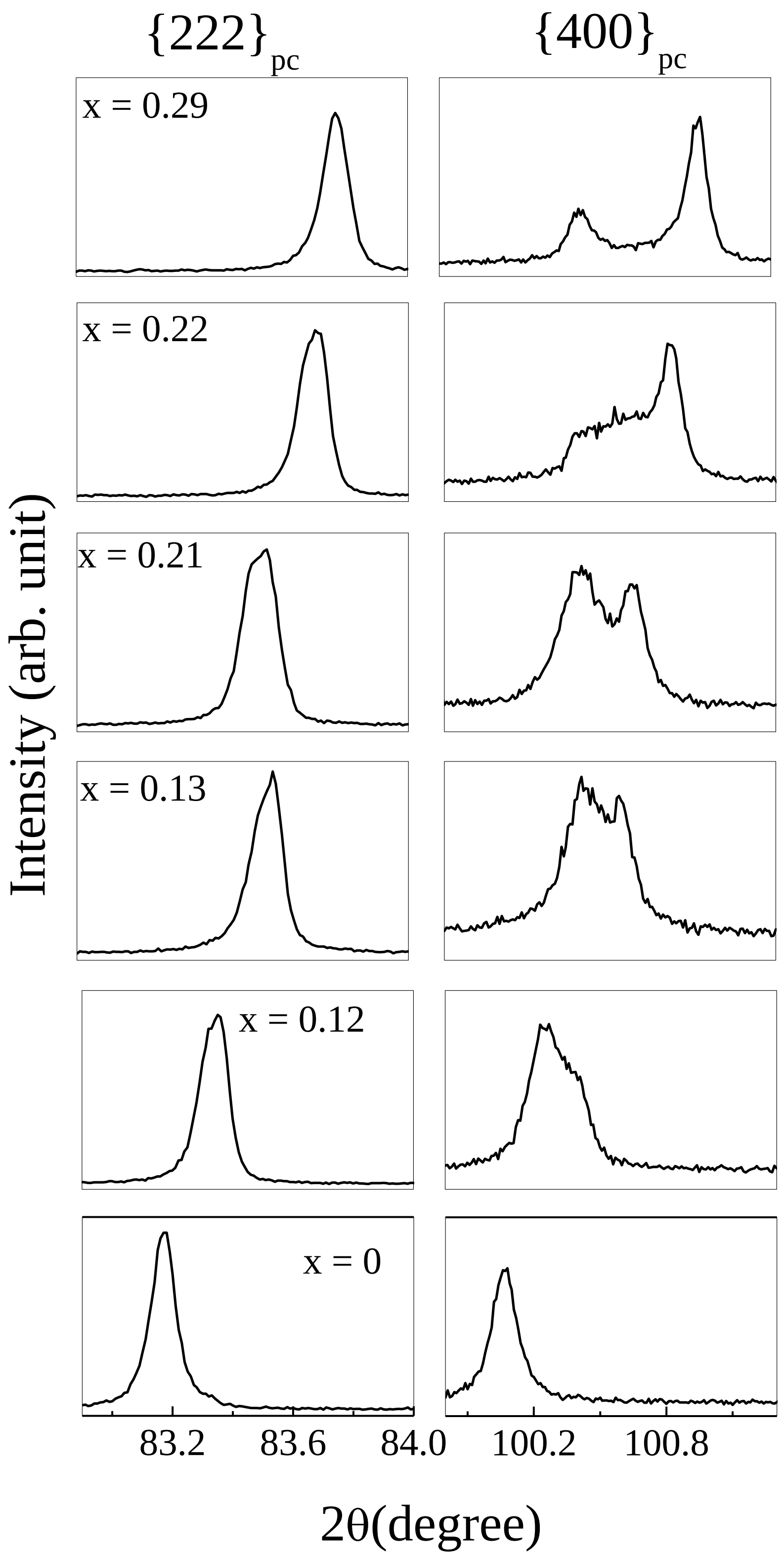

Fig. 2. X-ray Bragg profiles of the BCTS for x = 0, 0.12, 0.13, 0.21, 0.22, and 0.29.

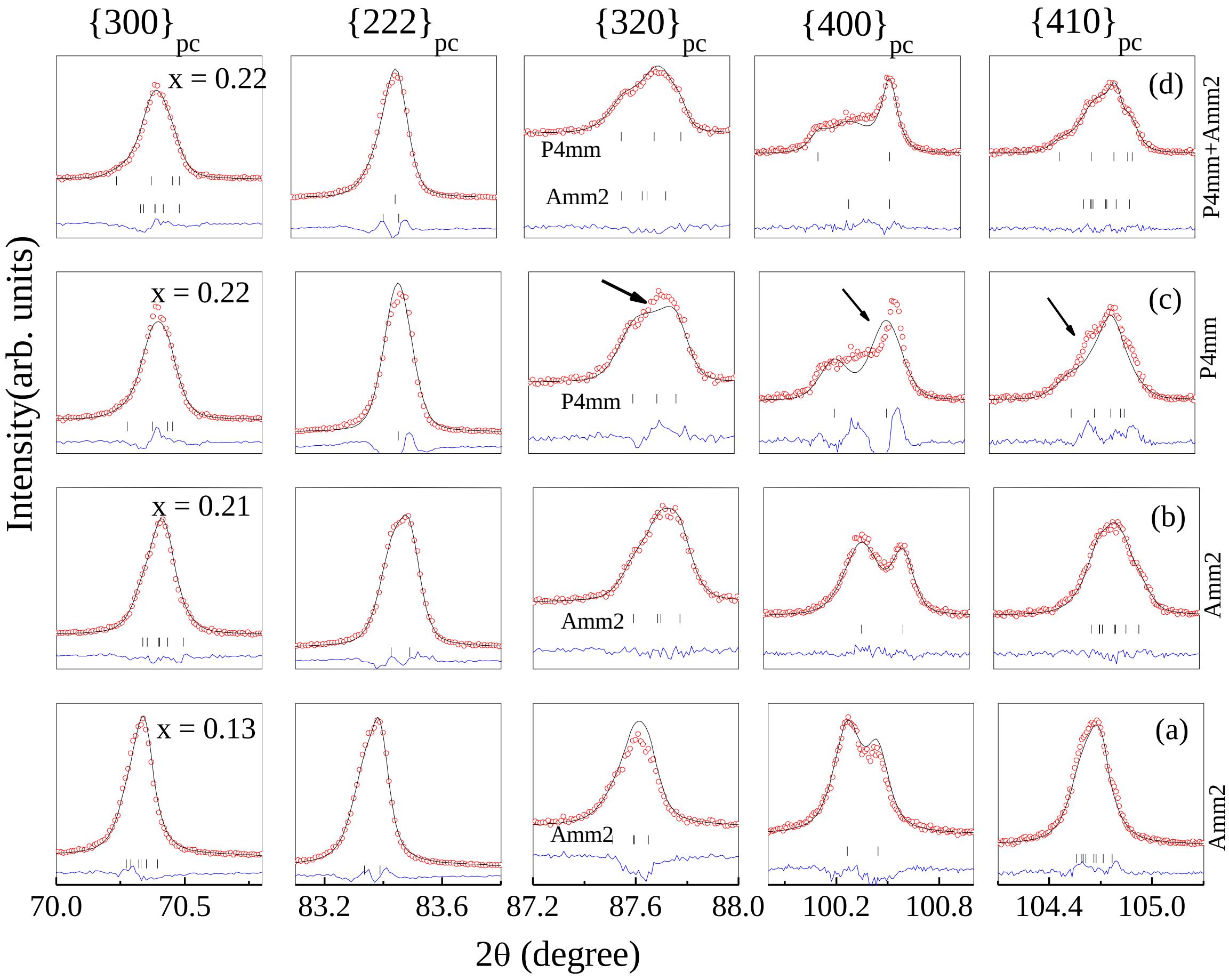

Fig. 3. Rietveld fitted x-ray powder differaction patternes of BCTS-100x for (a) x = 0.13 with Amm2, (b) x = 0.21 with Amm2, (c) x = 0.22 with P4mm, and (d) x = 0.22 with P4mm + Amm2. The arrows highlight the misfit regions.

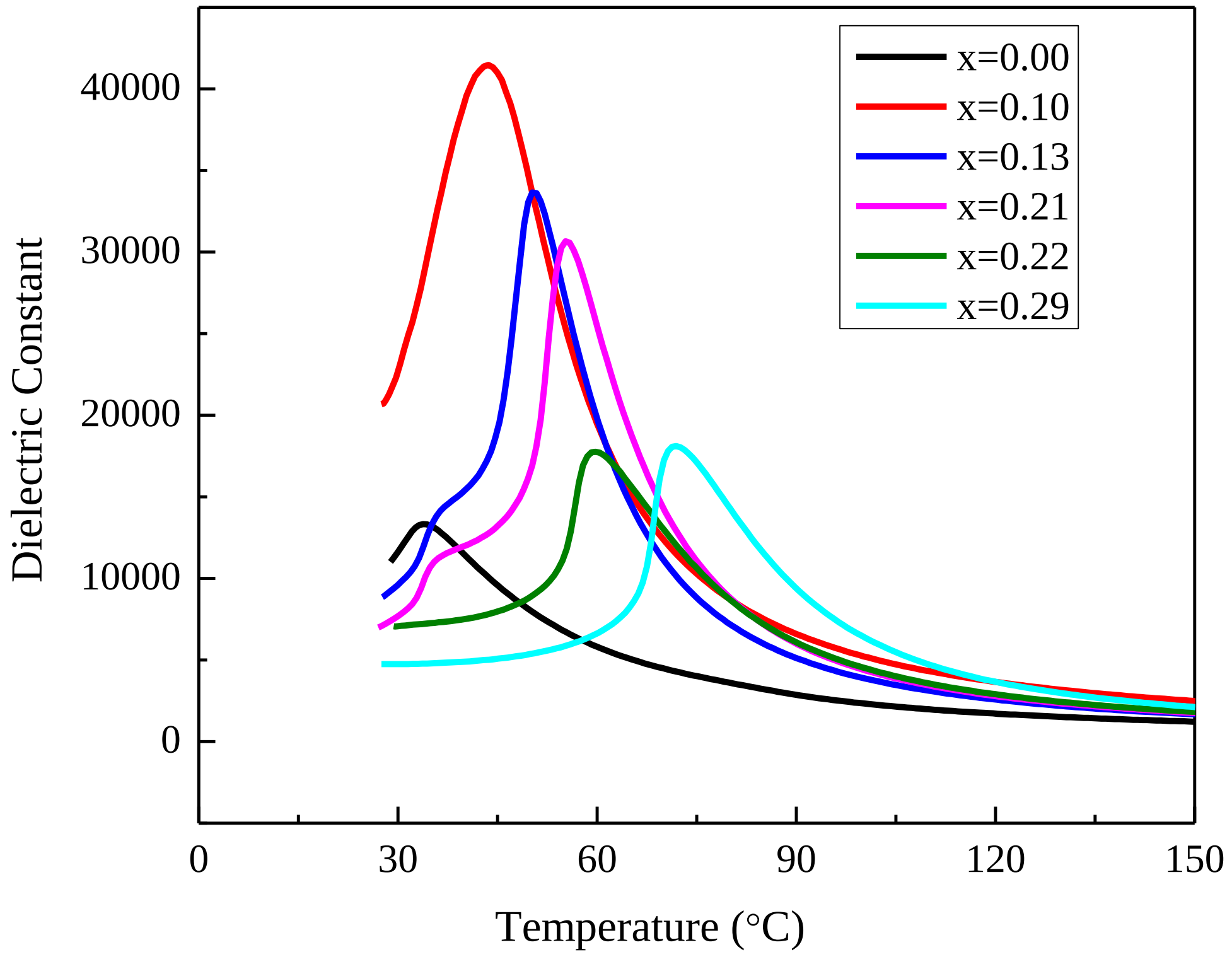

Fig. 4 Temperature dependence of the real part of relative permittivity of BCTS for x = 0, 0.10, 0.13, 0.21, 0.22, and 0.29 measured at 1 kHz.

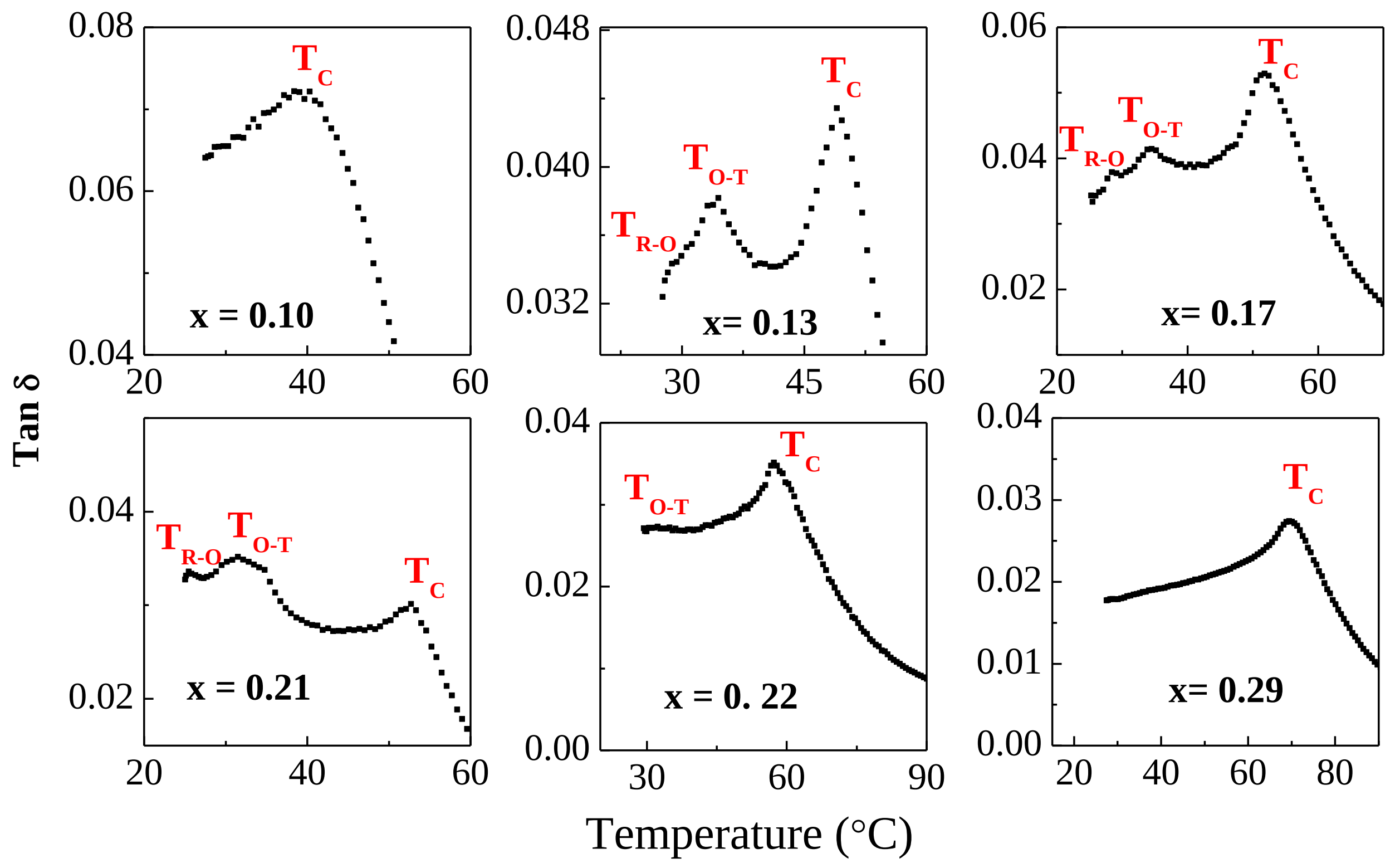

Fig. 5 Temperature dependent dielectric loss tangent of BCTS for x = 0.10, 0.13, 0.17, 0.21, 0.22, and 0.29 measured at 1 kHz.

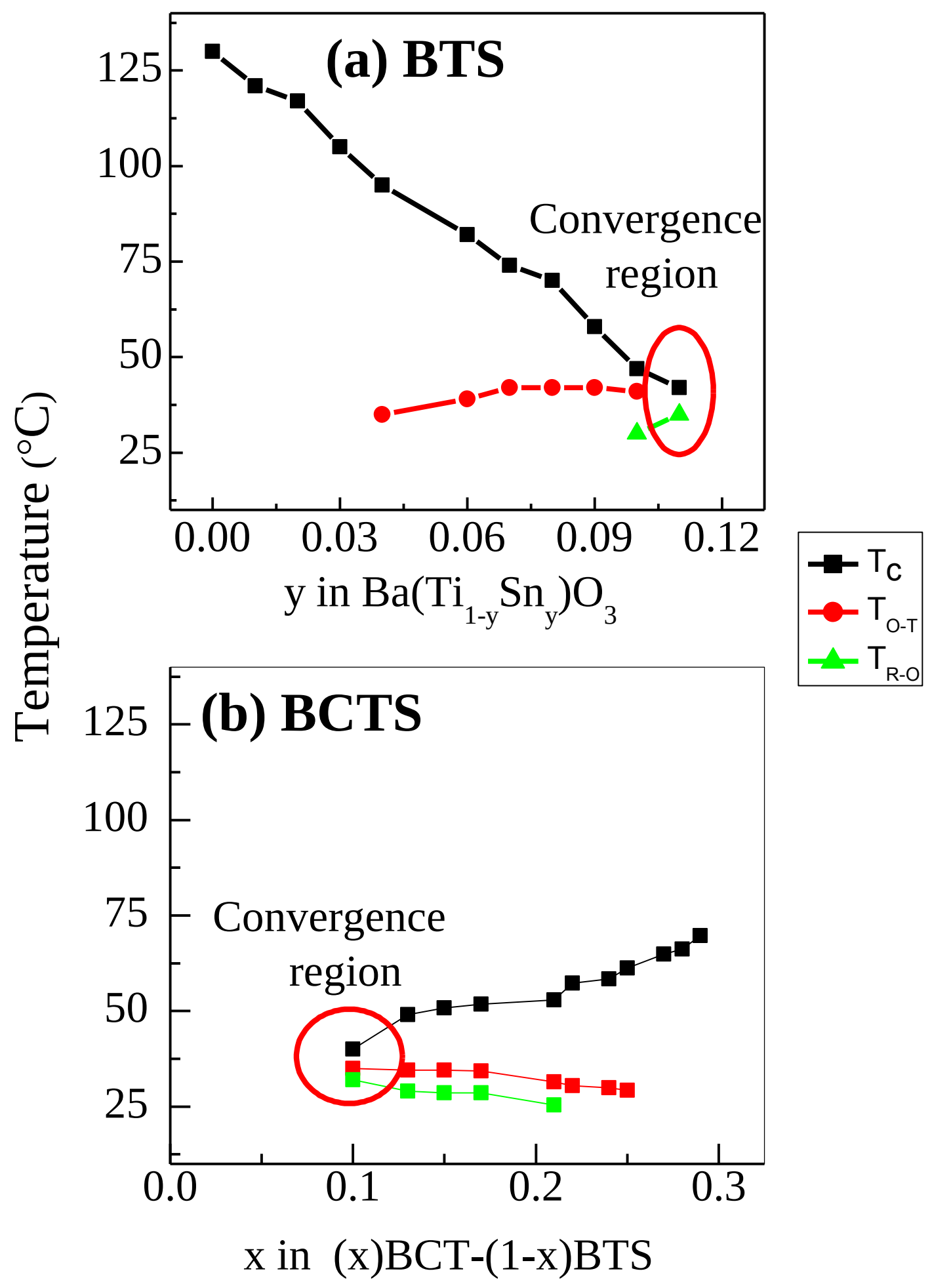

Fig. 6 (a) Phase diagram of (a) BTS – 100y and (b) (1-x)BCTS-100x ceramics on the basis of the dielectric anomalis observed in the imaginary part of the permittivity.

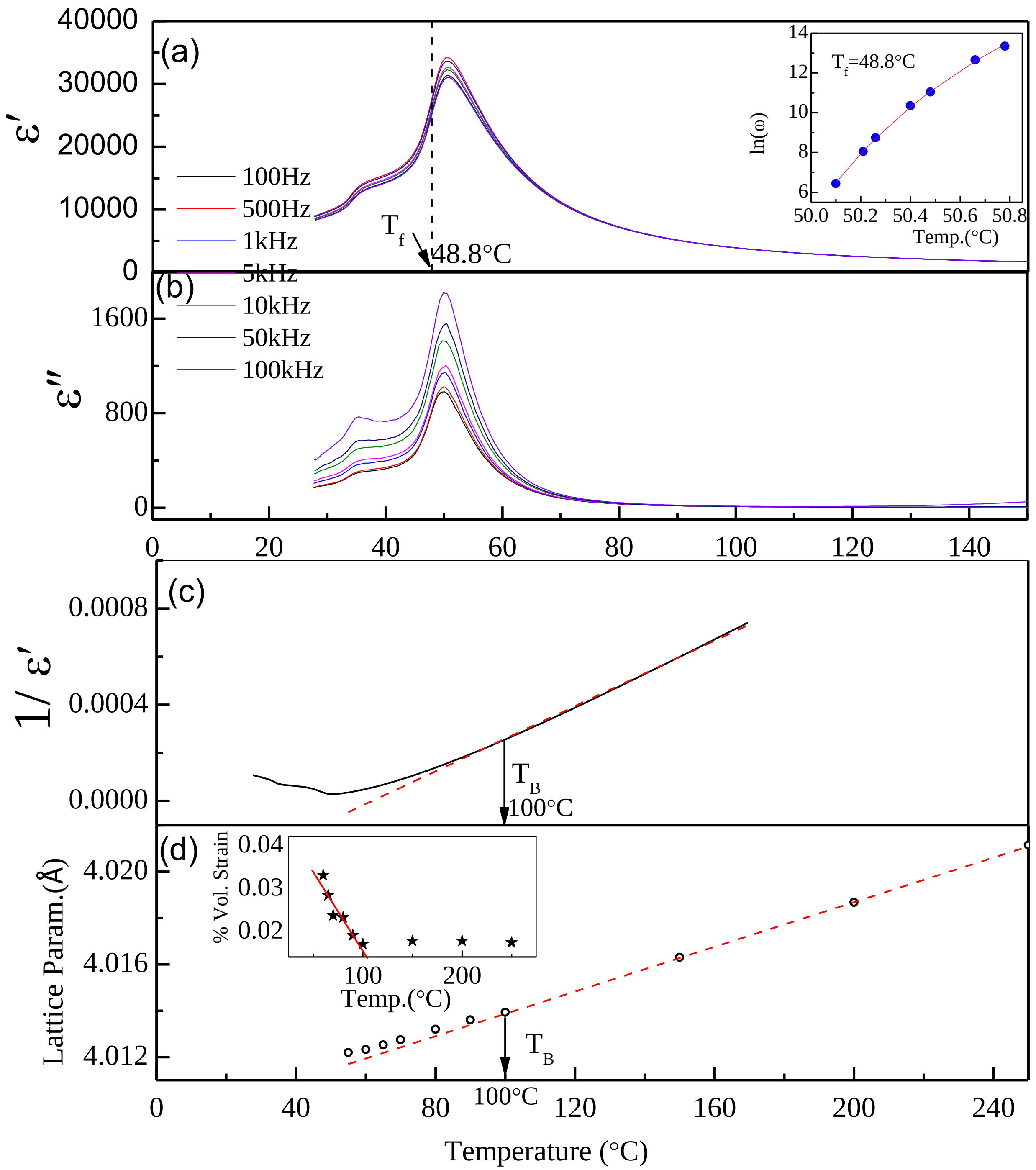

Fig. 7 Temperature variation of (a) real part and (b) imaginary part of the permittivity at different frequencies of BCTS with x =0.13. The inset in (a) shows the Vogel-Fulcher fit as per the equation $\omega = \omega_0 \exp[-E_a/k(T_m-T_f)]$, where $\omega_0$ (best fit value = 5.12 x $10^{10}$ Hz) is the theoretical maximum frequency for the vibration of $PNR_s$, $E_a$ (best fit value = 25 meV) is an activation energy, $k$ is the Boltzmann constant, $T_m$ is the temperature of the permittivity maximum at frequency $\omega$, and $T_f$ (best fit value = 48.8 $^0$C) is the freezing temperature that indicates the transition between the ergodic and the nonergodic states. (c) shows the Curie-Weiss fit of the real part of the permittivity data. $T_B$ corresponds to the Burn's temperature. (d) shows the variation of cubic lattice parameter with temperature. The inset in (d) shows percentage electrostrictive volume strain as a function of temperature below Burn's temperature.

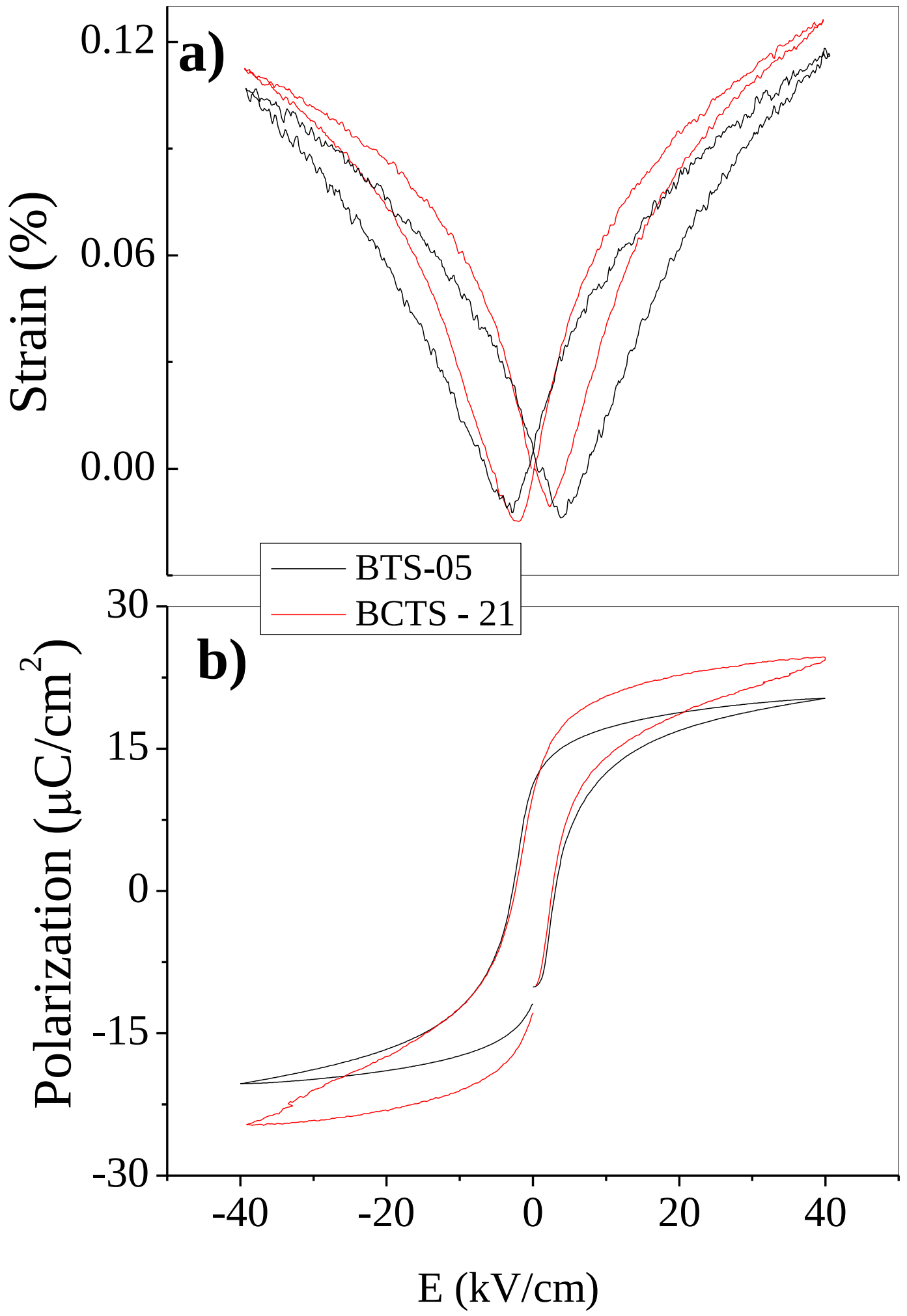

Fig.8. (a) polarization and (b) strain – field hysteresis loops of BTS-05 and BCTS-21.

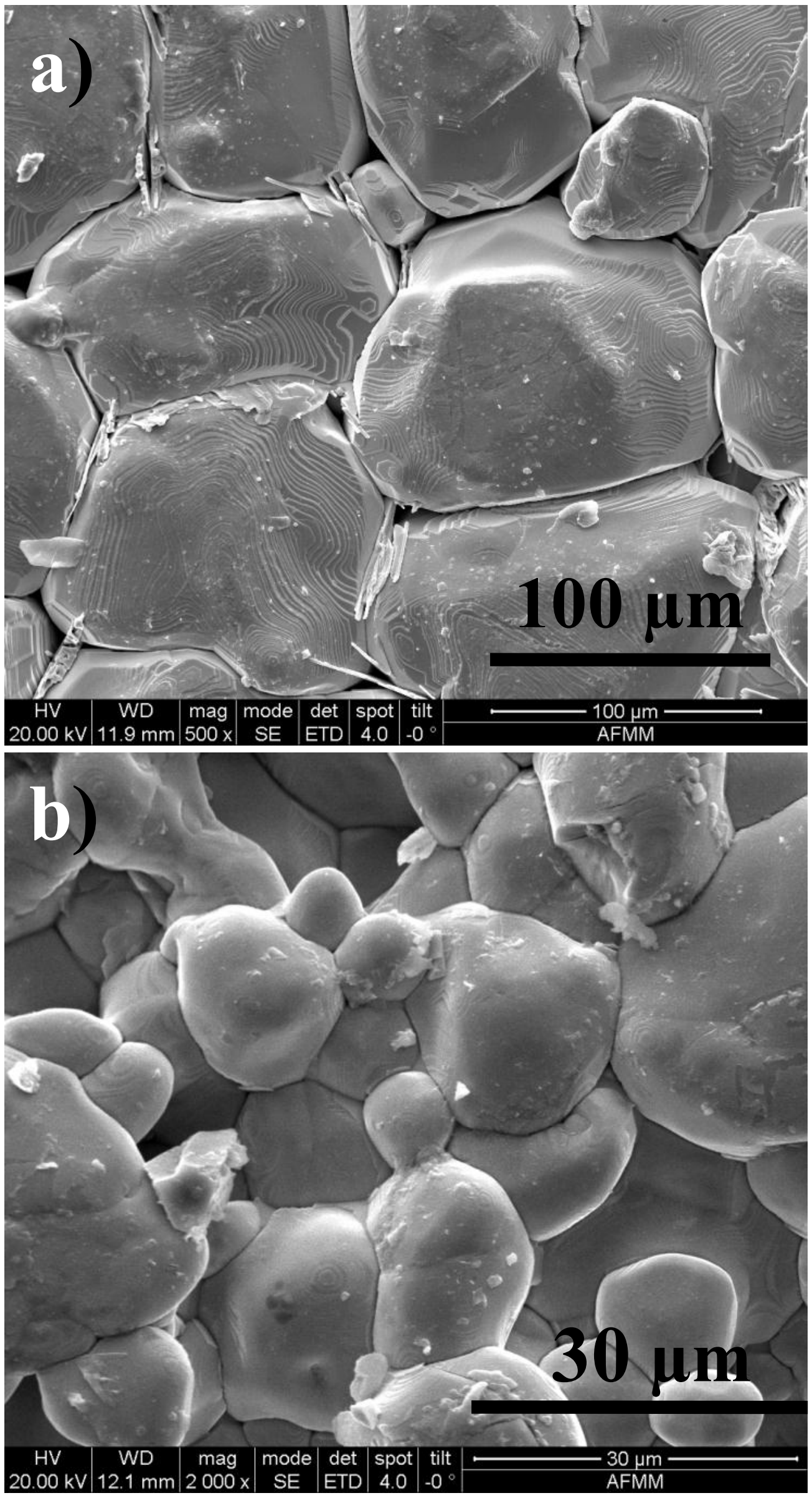

Fig. 9 SEM micrographs of (a) BTS-05 and b) BCTS-21.

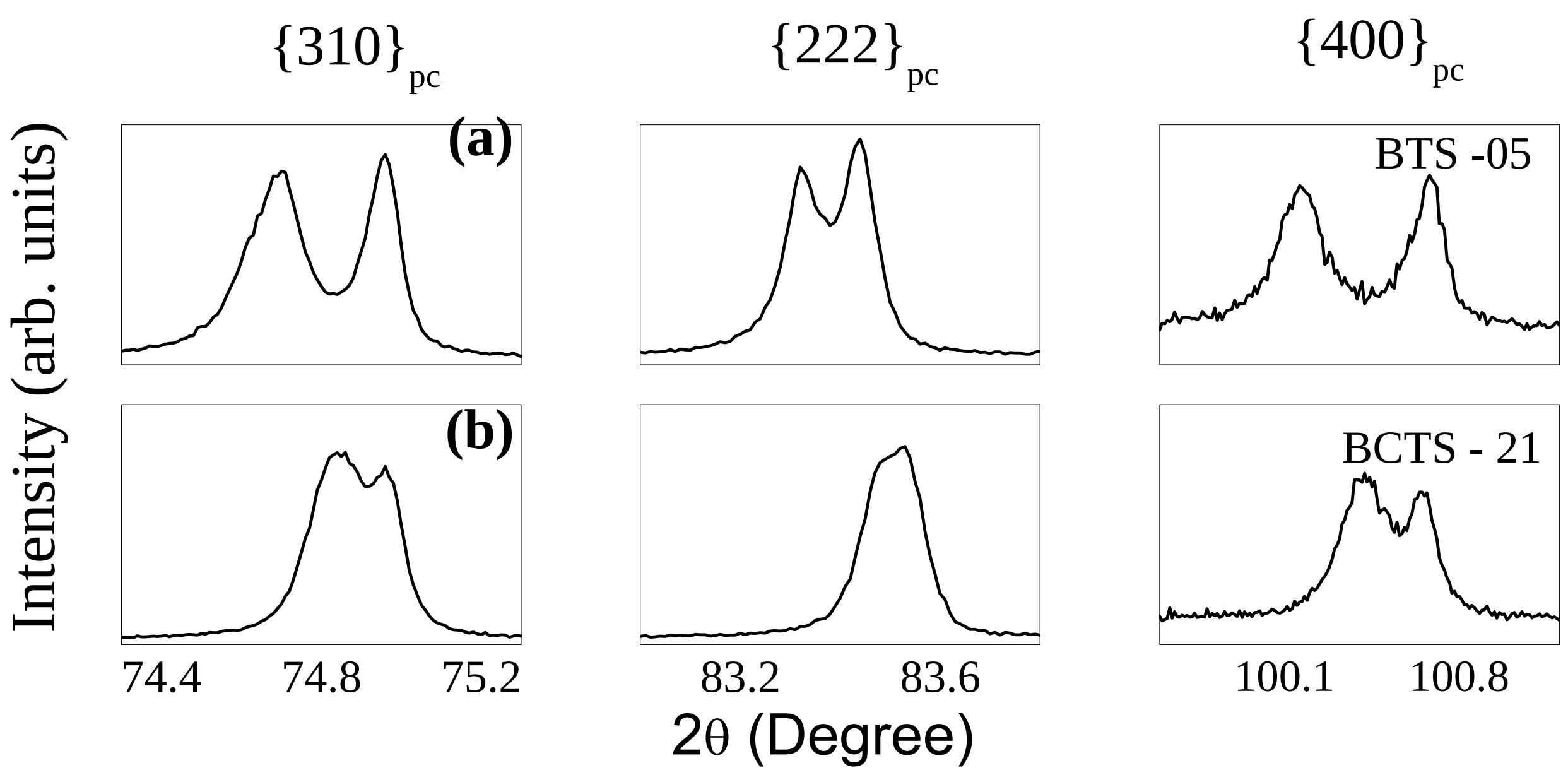

Fig. 10 A comparison of the x-ray powder diffraction Bragg profiles of selected pseudocubic peaks of (a) $BaTi_{0.95}Sn_{0.05}O_3$ (BTS-05) and (b) $0.79Ba(Ti_{0.88}Sn_{0.12})O_3 - 0.21(Ba_{0.7}Ca_{0.3})TiO_3$ (BCTS -21).

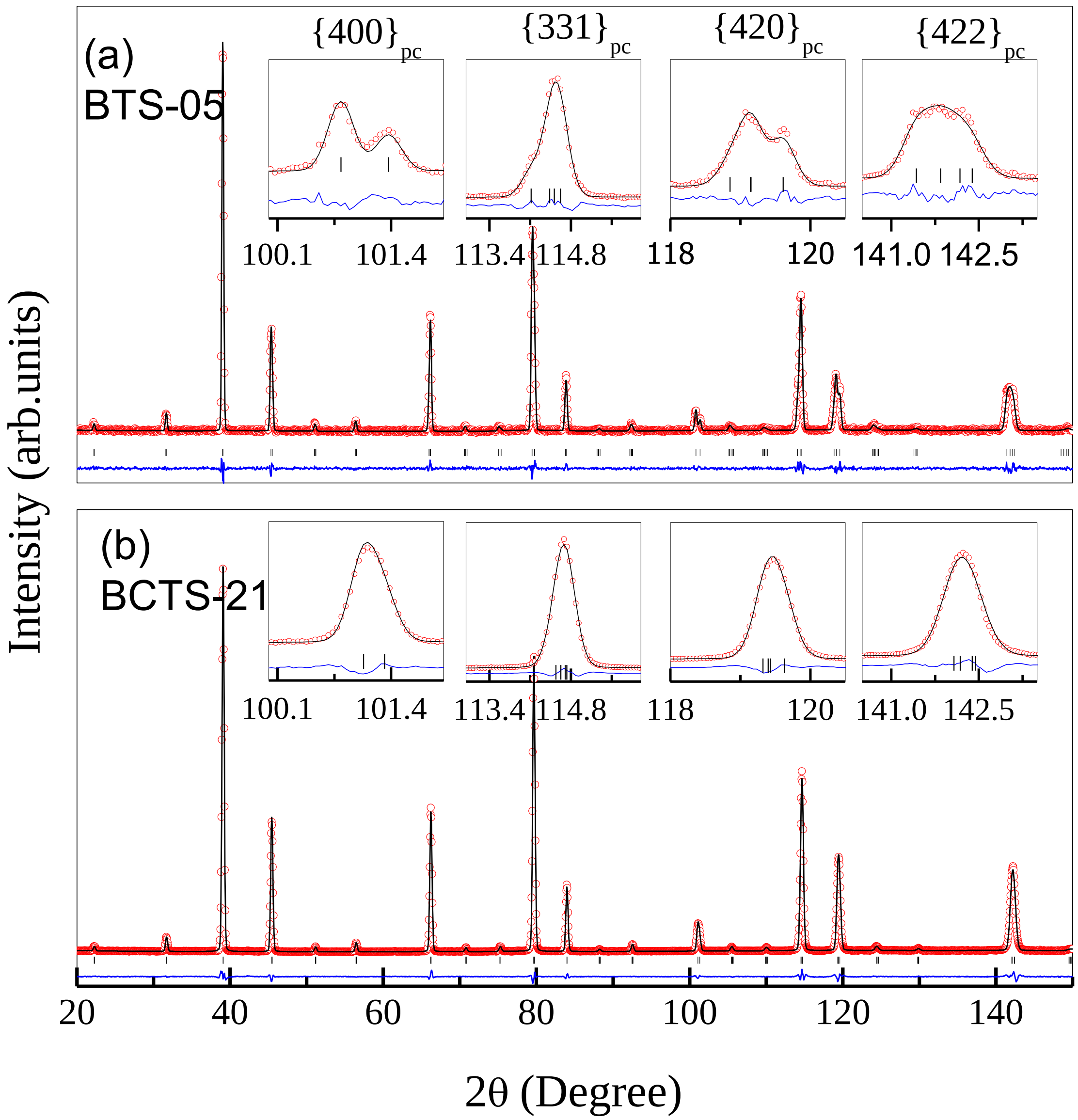

Fig 11. Rietveld fitted neutron powder diffraction pattern of (a) $BaTi_{0.95}Sn_{0.05}O_3$ (BTS - 05) and (b) $0.79Ba(Ti_{0.88}Sn_{0.12})O_3 – 0.21(Ba_{0.7}Ca_{0.3})TiO_3$ (BCTS - 21) with the Amm2 strucutral model.

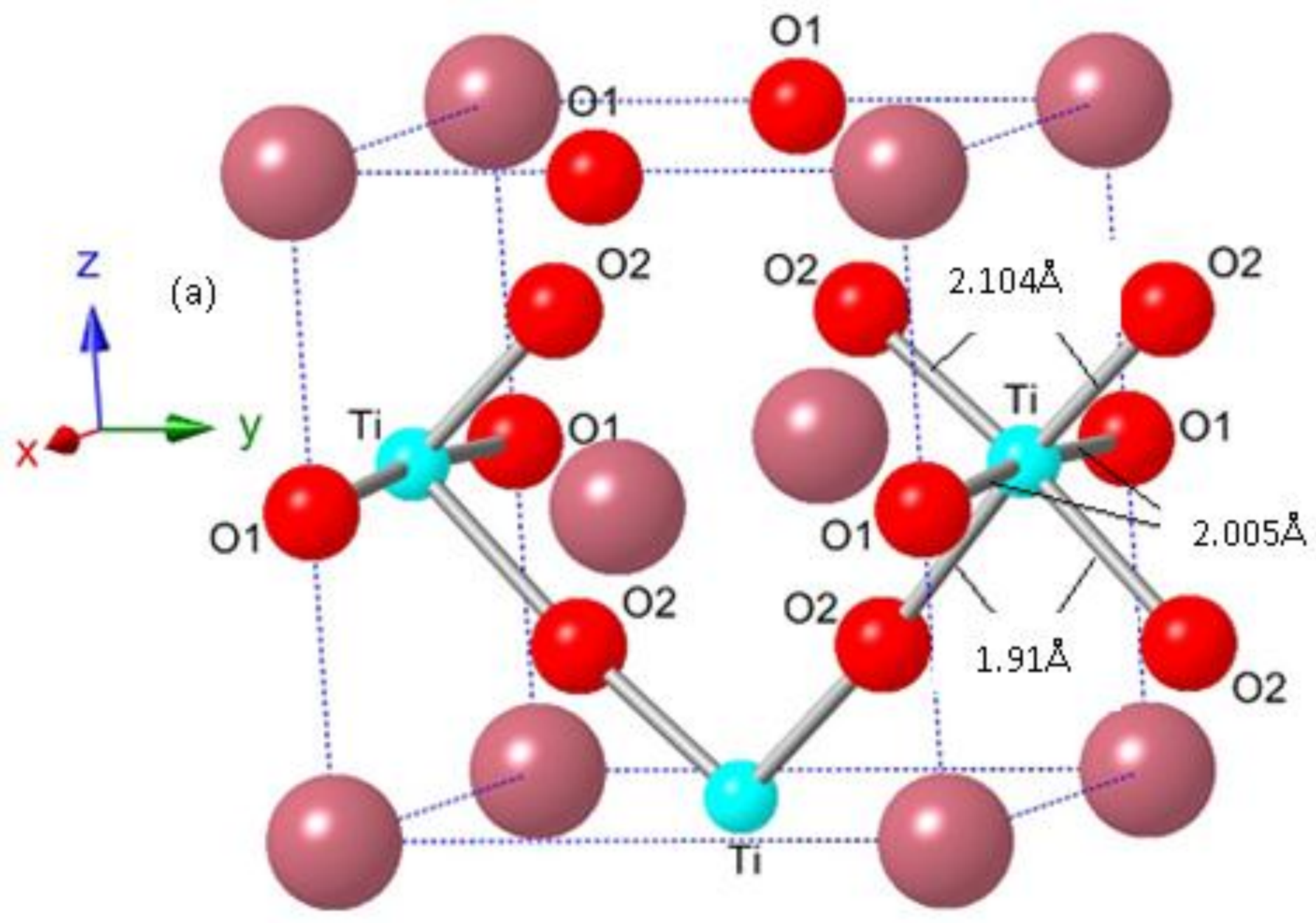

Fig. 12. Crystal structure of the orthorhombic phase of BCTS - 21. The large spheres in light maroon color represent Ba/Ca atoms and the medium spheres in red color O atoms. Small spheres in light blue color represent Ti/Sn atoms.